\documentstyle[epsf,epsfig,aps]{revtex}
\draft
\input epsf
\begin{document}
\twocolumn[\hsize\textwidth\columnwidth\hsize\csname@twocolumnfalse%
\endcsname
\title{Existence of the Abrikosov vortex state in two-dimensional type-II
superconductors without pinning.}
\author{A. \ V. \ Nikulov}

\address{Institute of Microelectronics Technology and High Purity Materials, Russian Academy of Sciences, 142432 Chernogolovka, Moscow District, RUSSIA}

\maketitle

\begin{abstract}
{ A nonperturbative fluctuation theory of the mixed state of
two-dimensional type-II superconductor with finite size in the lowest
Landau level approximation is proposed. The thermodynamic
averages of the spatial average order parameter and of the  Abrikosov
parameter $\beta_{a}$ are calculated. It is shown that the thermodynamic
fluctuations eliminate the Abrikosov vortex state in a wide region of the
mixed  state  of two-dimensional type-II superconductor with real
size. }  \end{abstract}
  ]

 \narrowtext

     Fluctuating behavior near second critical field, $H_{c2}$, in  type-II
superconductors has recently been under intense experimental
\cite{ber,yazd,kwok,marc,nik94} and theoretical
\cite{larkin,fish,maki89,moore89,lee,ikeda,houg,hikami,brezin,tes91,tes94,serg,nik90} 
study, particularly in connection with high temperature
superconductors (HTSC).  Investigation of the fluctuations changes step by
step our habitual notion about  nature of the mixed state. Concepts "vortex
lattice melting" and "vortex liquid" \cite{larkin,fish} have become very
popular on first stage of a stormy investigation of fluctuation effects in
HTSC. The vortex lattice melting transition is determined experimentally by
a change of resistive properties in a perpendicular magnetic field
\cite{ber,kwok}. This transition was observed in bulk
conventional superconductors before the HTSC discovery \cite{nik81a}. It
was interpreted in \cite{nik81a} as a transition from the Abrikosov state
into a one-dimensional state. But the vortex
liquid as well as the one-dimensional state is not a new genuine
thermodynamic state which is qualitatively different from the normal state
\cite{larkin}. Therefore the interpretations by \cite{ber,kwok} and
by \cite{nik81a} are just the same. The vortex lattice melting is the
transition from the Abrikosov state into the normal state with
superconducting fluctuations (which was called as the one-dimensional state
in \cite{nik81a,nik90}. The experimental
investigations show that this transition occurs below $H_{c2}$ in all case
\cite{nik81,nik84,ber,yazd,kwok}. The second critical field line
$H_{c2}(T)$ only marks a crossover from the normal metal to a strongly
fluctuating superconducting state \cite{larkin}. I think that the
widespread term "vortex lattice melting" is not enough right.
Therefore I will use the term "transition from the Abrikosov state into the
normal state" in exchange for it.

The starting point of vortex lattice melting theories is assumption of the
existence of the Abrikosov state at low temperatures. This assumption bases
oneself on experimental evidence only. But real samples with finite sizes
are investigated experimentally whereas the thermodynamic limit is used and
a influence of disorder (pinning centers) is not considered in the theory.
Therefore the experimental evidence may be not enough. The perturbation
fluctuation theory \cite{rugg76,brezin} shows no sign of the Abrikosov
state. Moreover, according to \cite{tes94} the fluctuation eliminates the
Abrikosov state in superconductors with infinite sizes.  The reduction of
effective dimensionality of the fluctuation in a magnetic field
\cite{lee72} and the destruction of off-diagonal long range order by
thermal fluctuations in the mixed state \cite{maki71,moore89,ikeda,serg}
cause doubt of the existence of the Abrikosov state in the thermodynamic
limit also.  According to results of \cite{maki71} the fluctuation can not
be considered as a perturbation in the mixed state of type-II
superconductor with infinite sizes. Therefore a nonperturbative
fluctuation theory of superconductors with finite sizes is required.

This theory for a most easy case is presented in this work. A
two-dimensional superconductor in a perpendicular magnetic field near
$H_{c2}$ is considered. In this case the lowest Landau level (LLL)
approximation is valid and the system is zero-dimensional \cite{tes91}.
Therefore the exact  thermodynamic averages can be calculated easily. The
theory is based on a new expression of the free energy obtained from the
Ginsburg-Landau (GL) one, $F_{GL}$. The results obtained in this work are
most close to one obtained in \cite{moore89}. In comparison with
\cite{moore89} a more easy method of calculation is suggested here and the
concept of a transition into the Abrikosov state of two-dimensional
superconductors with finite sizes is proposed.

We will proceed from the accurate expression for the free energy
$$F = -k_{B}T \ln\sum \exp(-\frac{F_{GL}}{k_{B}T}) \eqno{(1)}$$
Strongly type-II superconductors near $H_{c2}$ will be considered there. In
this case one can neglect the fluctuations in the magnetic  field.
The problem of fluctuations can be investigated within a framework  of
the GL free energy functional, with the order  parameter  confined  to
the lowest Landau level (LLL) for Cooper pairs. In this  approximation
the GL free energy functional for two-dimensional  superconductor  may
be write as (see \cite{tes91})

$$\frac{F_{GL}}{k_{B}T} = V(\varepsilon\overline{|\Psi|^{2}} +
0.5\beta_{a}\overline{|\Psi|^{2}}^{2}) \eqno{(2)}$$
$\varepsilon  = Gi^{-1/2}(ht)^{-1/2}(h-h_{c2})$, $Gi =
[2k_{B}T/H_{c}^{2}(0)d\xi^{2}(0)]$ is the  Ginsburg number in zero
magnetic field; $t = T/T_{c}$; $h = H/H_{c2}(0)$;  $H_{c2}(0) =
-T_{c}(dH_{c2}/dT)_{T=T_{c}}$; $H_{c}(0) = - T_{c}(dH_{c}/dT)$; $H_{c}$
is the thermodynamic critical field; $\xi (0)$ is the coherence length at T
= 0. d is  the  film thickness; V is the film area; $\overline{|\Psi|^{2}}
= (\int_{V} d^{2}r|\Psi|^{2})/V$ is the  spatial average order parameter;
$\beta_{a} = \overline{|\Psi|^{4}}/\overline{|\Psi|^{2}}^{2}$ is the
generalized  Abrikosov parameter; $\overline{|\Psi|^{4}} = (\int_{V}
d^{2}r|\Psi|^{4})/V$.  We use a  dimensionless  unit system in  which
$l_{x} = 1$; $l_{y}/2 = 1$; $l_{x}l_{y}/2 = \Phi_{0}/H = 1$; $\Psi =
(d\beta l_{x}l_{y}/2k_{B}T)^{1/4}\psi$. In this unit system the film area
V  is equal  the degeneracy number of the LLL. $l_{x}, l_{y}$  are the
parameter of the triangular Abrikosov lattice corresponding to a magnetic
field value H; $\Phi_{0}$ is the flux quantum; $\psi$ is order parameter in
the conventional units.  To obtain (2) we used the well-known relation:
$\alpha = -e\hbar H_{c2}/mc$;  $\alpha^{2}/2\beta = H_{c}^{2}/8\pi$.
$\alpha$ and $\beta$ are the conventional coefficients of  GL  free  energy
\cite{genn}.

In the LLL approximation the sum in (1) is to be taken  over  the
subspace spanned by the LLL. To substitute the sum in (1) on two integral

$$F = -k_{B}T \ln[\int_{0}^{\infty} d\overline{|\Psi|^{2}}
\int_{\beta_{A}}^{\infty} d\beta_{a} N(\overline{|\Psi|^{2}};\beta_{a})
 \exp(-\frac{F_{GL}}{k_{B}T)}] \eqno{(1a)}$$
$N(\overline{|\Psi|^{2}};\beta_{a})d\overline{|\Psi|^{2}}d\beta_{a}$ is
a subspace volume with the given  values  of  $\overline{|\Psi|^{2}}$
and $\beta_{a}$. $\beta_{A}$ is the minimum value of $\beta_{a}$ in the
subspace spanned by LLL. It was shown in \cite{klein} that $\beta_{A}$
correspond to the triangular lattice and is equal approximately
1.159595.

     To expand the order parameter on the Landau eigenfunctions $\Psi (r) =
V^{-1/2}\sum_{\lambda} \Psi_{\lambda}\varphi_{\lambda}(r)$. Then
$\overline{|\Psi|^{2}} = V^{-1} \sum_{\lambda} |\Psi_{\lambda}|^{2} =
V^{-1} \sum_{\lambda} [(Re\Psi_{\lambda})^{2} + (Im\Psi_{\lambda})^{2}]$.
A eigenfunction number is equal the  degeneracy  number  of  the  LLL,
which is equal in our unit system V. Consequently the given values  of
$\overline{|\Psi|^{2}}$ lie on a 2V-dimensional sphere with radius
$\overline{|\Psi|^{2}}^{0.5}$: $(\overline{|\Psi|^{2}}^{0.5})^{2} =
\sum_{\lambda} [(Re\Psi_{\lambda}/V^{0.5})^{2} +
(Im\Psi_{\lambda}/V^{0.5})^{2}]$. The  area  of
this  sphere  is equal $2\pi^{V} V!
(\overline{|\Psi|^{2}}^{0.5})^{2V-1}$.  It is obvious that $\beta_{a}$
value does not depend on the $\overline{|\Psi|^{2}}$ value. Therefore
$N(\overline{|\Psi|^{2}};\beta_{a})$ is proportional to
$\overline{|\Psi|^{2}}^{V-0.5} n(\beta_{a})$. $n(\beta_{a})$ is a
fraction  of  the  2V-dimensional  sphere with the given value of
$\beta_{a}$. Multipliers independent of $\overline{|\Psi|^{2}}$ and
$\beta_{a}$ are omitted.

To write the generalized Abrikosov parameter through  eigenvalues
$\Psi_{\lambda}$, $\beta_{a} = (\sum_{\lambda_{i}}
V_{\lambda_{1},\lambda_{2},\lambda_{3},\lambda_{4}} \Psi_{\lambda_{1}}^{*}
\Psi_{\lambda_{2}}^{*} \Psi_{\lambda_{3}}
\Psi_{\lambda_{4}})/(\sum_{\lambda} |\Psi_{\lambda}|^{2})^{2}$.
$V_{\lambda_{1},\lambda_{2},\lambda_{3},\lambda_{4}} = V^{-1} \int_{V}
d^{2}r \varphi_{\lambda_{1}}^{*} \varphi_{\lambda_{2}}^{*}
\varphi_{\lambda_{3}} \varphi_{\lambda_{4}}$. We shall use the Eilenberger
basis function \cite{eilen}. In this basis each $\varphi_{\lambda}$
function describes the triangular Abrikosov lattice and $\lambda =
(\lambda_{x};\lambda_{y})$ is their relative displacement. $\lambda_{x} =
(l_{x}^{2}/L_{y})n$;  $\lambda_{y} = (l_{y}^{2}/L_{x})m$. n; m are
integer numbers; $L_{x}; L_{y}$ are film sizes. $L_{x}L_{y} = V$.
$V_{\lambda_{1},\lambda_{2},\lambda_{3},\lambda_{4}}$ was calculated in
\cite{rugg}.

Let $\Psi_{\lambda=0} = \Psi_{0} = 1$. Following  \cite{maki71}
I introduce  new  variables  $u_{+}(\lambda)$ and $u_{-}(\lambda)$

$$\Psi_{\lambda} =
(|V_{0,0,\lambda,-\lambda}|/2V_{0,0,\lambda,-\lambda})^{1/2} [u_{+}(\lambda
) + u_{-}(\lambda )]$$

$$\Psi_{-\lambda} =
(|V_{0,0,\lambda,-\lambda}|/2V_{0,0,\lambda,-\lambda})^{1/2}
[u_{+}^{*}(\lambda ) - u_{-}^{*}(\lambda )]$$

Then the generalized Abrikosov parameter may be write as

$$\beta_{a} = \beta_{A} + \frac{\sum_{\lambda \neq 0} (\Gamma_{+}
|u_{+}|^{2} + \Gamma_{-} |u_{-}|^{2}) + U - B}{[1 + 0.5\sum_{\lambda \neq
0} (|u_{+}|^{2} + |u_{-}|^{2})]^{2}} \eqno{(3a)}$$

$$\Gamma_{+} = 2V_{0,\lambda,0,\lambda} + |V_{0,0,\lambda,-\lambda}| -
\beta_{A}$$   $$ \Gamma_{-} = 2V_{0,\lambda,0,\lambda} -
|V_{0,0,\lambda,-\lambda}| - \beta_{A}$$

$$U = \sum_{\lambda_{i} \neq 0}
V_{\lambda_{1},\lambda_{2},\lambda_{3},\lambda_{4}} \Psi_{\lambda_{1}}^{*}
\Psi_{\lambda_{2}}^{*} \Psi_{\lambda_{3}} \Psi_{\lambda_{4}}$$     $$B =
\beta_{A}(\sum_{\lambda \neq 0} |\Psi_{\lambda}|^{2})^{2}$$
The sum by all $\lambda \neq 0$.

At $B \ll 1$, $\beta_{a} \simeq \beta_{A} + \sum_{\lambda \neq 0}
  (\Gamma_{+} |u_{+}|^{2} + \Gamma_{-} |u_{-}|^{2})$. Consequently at
$\beta_{a} - \beta_{A} \ll 1$ the given value $\beta_{a}$ lie  in  a
2V-dimensional  circle.  A "length" of this circle is proportional to
$[(\beta_{a} - \beta_{A})^{0.5}]^{2V-2}$. Therefore $n(\beta_{a})$ is
proportional to $(\beta_{a} - \beta_{A})^{V-1}$ at $\beta_{a} - \beta_{A}
\ll 1$. At a larger  $\beta_{a} - \beta_{A}$ value $n(\beta_{a})$ increases
more slowly with the $\beta_{a} - \beta_{A}$ increasing.  In  this  case
one may introduce a new function and write $n(\beta_{a}) = [f(\beta_{a} -
\beta_{A})]^{V-1}$.  Taking into account that $V \gg 1$ we may overwrite
(1a) as

$$F = -k_{B}T \ln[\int_{0}^{\infty} d\overline{|\Psi|^{2}}
\int_{\beta_{A}}^{\infty} d\beta_{a}
\exp(-\frac{VF_{New}}{k_{B}T})] \eqno{(1b)}$$
where $F_{New}$ is the new free energy expression in which the entropy term
connected with $\overline{|\Psi|^{2}}$ and $\beta_{a}$ values is take into
account
$$\frac{F_{New}}{k_{B}T} = \varepsilon \overline{|\Psi|^{2}} +
0.5\beta_{a}\overline{|\Psi|^{2}}^{2} - \ln\overline{|\Psi|^{2}} -
\ln[f(\beta_{a} - \beta_{A})] \eqno{(4)}$$
Because $V \gg 1$ the exact thermodynamic averages of the
$\overline{|\Psi|^{2}}$ and $\beta_{a}$ values, $<\overline{|\Psi|^{2}}>$
and $<\beta_{a}>$, are close to values corresponded to the $F_{New}$
minimum.  Consequently these values are specified by relations

$$<\overline{|\Psi|^{2}}> = [(\frac{\varepsilon}{2<\beta_{a}>})^{2}
+ \frac{1}{<\beta_{a}>}]^{1/2} - \frac{\varepsilon}{2<\beta_{a}>}
\eqno{(5a)}$$

$$\frac{f'(<\beta_{a}> - \beta_{A})}{f(<\beta_{a}> - \beta_{A})} =
0.5<\overline{|\Psi|^{2}}>^{2} \eqno{(5b)}$$
The (5a) relation coincide with one obtained in \cite{tes94} and is close
to the entropy dependence obtained by O'Neill and Moore \cite{moore89}.
This relation is close to the Abrikosov solution, $<\overline{|\Psi|^{2}}>
= -\varepsilon /<\beta_{a}>$, at $-\varepsilon \gg 1$ (below the $H_{c2}$
critical region) and to the result of the  fluctuation theory in the linear
approximation, $<\overline{|\Psi|^{2}}> = 1/\varepsilon$, at $\varepsilon
\gg 1$ (above the $H_{c2}$ critical region).

The $f(\beta_{a} - \beta_{A})$ function is not evaluated completely in this
work.  Therefore we calculate the thermodynamic averages of $\beta_{a}$
below the critical region only. At $-\varepsilon \gg 1$, $<\beta_{a}> -
\beta_{A} \ll 1$. Consequently $f(<\beta_{a}> - \beta_{A}) \simeq
<\beta_{a}> - \beta_{A}$ and according to (5b)

$$<\beta_{a}> - \beta_{A} = 2<\overline{|\Psi|^{2}}>^{-2} \simeq
2(\frac{\beta_{A}}{\varepsilon})^{2} = $$
$$2\beta_{A}^{2} Gi ht(h_{c2} - h)^{-2}
\eqno{(6)}$$

 The Ginsburg number values of conventional two-dimensional type-II
 superconductors lie in the interval $3*10^{-6} < Gi < 3*10^{-3}$.
 Therefore in the center of the mixed state region, at t = 0.5, h = 0.25,
$10^{-5} < (<\beta_{a}> - \beta_{A}) < 10^{-2}$.

According (1b) there is not thermodynamic singularity  in
two-dimensional type-II superconductor which may be connected with a phase
transition from normal state into Abrikosov vortex state.  Therefore one
should think that this transition occurs without thermodynamic
singularity. I propose there a concept of this transition based  on  the
difference between the Abrikosov state and the normal state with
superconducting fluctuation in the Eilenberger function basis
\cite{eilen}.  It  is obvious that above the $H_{c2}$ critical region
all thermodynamic average of the Eilenberger  eigenvalue squares,
$<|\Psi_{\lambda}|^{2}>$, are equal and $<|\Psi(r)|^{2}>$ is constant in
space.  Consequently, in the normal state with superconducting
fluctuation (in the vortex liquid phase \cite{larkin})
$<\overline{|\Psi|^{2}}> = V^{-1} \sum_{\lambda} <|\Psi_{\lambda}|^{2}> =
<|\Psi_{\lambda=0}|^{2}>$ whereas in the Abrikosov state
$<\overline{|\Psi|^{2}}> = V^{-1}<|\Psi_{\lambda=0}|^{2}>$  in the mean
field approximation.  Therefore  one should think that the transition into
the Abrikosov vortex state  take place when a one Eilenberger eigenvalue,
$\Psi_{0}$ , brings the main contribution to the order parameter:
$<|\psi_{0}|^{2}> = <|\Psi_{0}|^{2}>/V<\overline{|\Psi|^{2}}>$ = $(1 +
\sum_{\lambda \neq 0} <|\psi_{\lambda}|^{2}>)^{-1}$ = $(1+0.5\sum_{\lambda
\neq 0} <|u_{+}|^{2}> + <|u_{-}|^{2}>)^{-1} \simeq 1$.  Where
$<|\psi_{\lambda}|^{2}> = <|\Psi_{\lambda}|^{2}>/<|\Psi_{0}|^{2}>$.

The $<|\psi_{0}|^{2}>$ value is connected with $<\beta_{a}> - \beta_{A}$
value.  We shall consider the region below the $H_{c2}$ critical region
where $<\beta_{a}> - \beta_{A} \ll 1$.  Following to \cite{maki71} one
may show that $\Gamma_{+} \simeq 2\beta_{A}$ and $\Gamma_{-} \simeq
3.1|\lambda|^{4}$ at $|\lambda| \ll 1$.  Consequently, according (3a),
$\sum_{\lambda \neq 0} |u_{+}|^{2} \ll 1$ and it may be that $|u_{-}|^{2}
\gg 1$ at $\beta_{a} - \beta_{A} \ll 1$.  Therefore we will consider
states with $u_{+} = 0$ only.  In this case $|\psi_{0}|^{2} = (1 +
0.5\sum_{\lambda \neq 0}|u_{-}|^{2})^{-1}$ and

$$\beta_{a} - \beta_{A} = \frac{\sum_{\lambda \neq 0} \Gamma_{-}
|u_{-}|^{2} + U -B}{(1 + 0.5\sum_{\lambda \neq 0}|u_{-}|^{2})^{2}}
\eqno{(3b)}$$
$U = 0.25\sum_{\lambda_{i} \neq 0}
V_{\lambda_{1},\lambda_{2},\lambda_{3},\lambda_{4}}
\exp[0.5i(\varphi_{\lambda_{1}} +\varphi_{\lambda_{2}}
-\varphi_{\lambda_{3}} - \varphi_{\lambda_{4}})]
u_{-}^{*}(\lambda_{1})u_{-}^{*}(\lambda_{2})u_{-}(\lambda_{3})
u_{-}(\lambda_{4})$; $B = \beta_{A}(0.5\sum_{\lambda \neq
0}|u_{-}|^{2})^{2}$.  Where $\exp(-i\varphi_{\lambda}) =
|V_{0,0,\lambda,-\lambda}|/V_{0,0,\lambda,-\lambda}$.

At $0.5\sum_{\lambda \neq 0}|u_{-}|^{2} \ll 1$ we have the Abrikosov state
with fluctuation and at $0.5\sum_{\lambda \neq 0}|u_{-}|^{2} \gg 1$ we have
the normal state with superconducting fluctuations.  All states
corresponding to very small $\beta_{a} - \beta_{A}$ value lie near the
Abrikosov state ($0.5\sum_{\lambda \neq 0}|u_{-}|^{2} \ll 1$). At
larger $\beta_{a} -\beta_{A}$ value states exist which are removed from the
Abrikosov state (states with $0.5\sum_{\lambda \neq 0}|u_{-}|^{2} \gg 1$).
The critical value, $(\beta_{a} - \beta_{A})_{c}$, exists below which
states with $0.5\sum_{\lambda \neq 0}|u_{-}|^{2} \gg 1$ appear.  I suppose
that transition into Abrikosov state occurs when $<\beta_{a}> - \beta_{A} =
(\beta_{a} - \beta_{A})_{c}$.

The $(\beta_{a} - \beta_{A})_{c}$ value is not calculated in this paper.  I
will show only that in two-dimensional superconductor states with
$0.5\sum_{\lambda \neq 0}|u_{-}|^{2} \gg 1$ exist at small $\beta_{a} -
\beta_{A}$ value.

We will consider the states with $u_{-} (\lambda_{x}=0;\lambda_{y}) =
u/(\lambda_{y}L_{x}) = u/m$ at $|\lambda_{y}| \leq l \leq 0.1$, m is odd
number and $u_{-}(\lambda_{x} ;\lambda_{y}) = 0$ at all other
$\lambda_{x},\lambda_{y}$ values.  In this case

$$\beta_{a} - \beta_{A} = (\frac{0.52 l^{3} u^{2} + (0.87/l - 2.8
l)u^{4}}{(1 + 1.23u^{2})^{2}}) \frac{1}{L_{x}} \eqno{(3c)}$$

At $0.5\sum_{\lambda \neq 0}|u_{-}|^{2} = 1.23u^{2} \gg 1$, $\beta_{a} -
\beta_{A} \simeq (0.58/l - 1.86l)/L_{x}$.  The relation (3c) is valid at
$l \leq 0.1$ only.  At l = 0.1, $\beta_{a} - \beta_{A} = 5.6/L_{x}$.
Consequently $(\beta_{a} - \beta_{A})_{c} \leq 5.6/L_{x}$.

Thus the obtained results show that the Abrikosov vortex state can not
exist in two-dimensional type-II superconductors with infinite sizes (at
$L_{x} = \infty, (\beta_{a} - \beta_{A})_{c} = 0$ and consequently
$<\beta_{a}> - \beta_{A} > (\beta_{a} - \beta_{A})_{c}$ always). The same
result was obtained in \cite{moore89,tes94}.

In "dirty" ($Gi \simeq 3 \ 10^{-3}$ ) thin films with usual sizes ($L_{x} =
L_{x}/l_{x} = 10^{3} - 10^{5}$) the transition into Abrikosov state takes
place much below $H_{c2}$ (for example at t = 0.5 and $Gi = 3*10^{-3}$,
$<\beta_{a}> - \beta_{A} < (\beta_{a} - \beta_{A})_{c} \leq 5.6/L_{x} =
5*10^{-3} - 5* 10^{-5}$ below $h = 0.25 \simeq 0.5h_{c2}$ only, see above).
It must be emphasize that these conclusion are valid for samples without
pinning centers only.  Any pinning of the flux lines by disorder, ets.,
will inhibit and slow the fluctuations and will stabilize the Abrikosov
state \cite{moore89}.

According to the concept described above, the Abrikosov
state and the normal state with superconducting fluctuation have different
distributions of the order parameter in the space.  Therefore a transition
from the Abrikosov state into the normal state (vortex lattice melting)
can be observed experimentally by pinning appearance (by a change of
resistive properties in a perpendicular magnetic field).  It was observed
in \cite{ber,nik94} that this transition occurs (pinning appears) much
below $H_{c2}$ in thin amorphous films with a small amount of pinning
centers. It may be considered an experimental confirmation of the result
obtained above.

I would like to thank Z.Tesanovic` for sending his paper.  
The results of these papers stimulated this work. This work was
supported by the National Scientific Council on High Temperature
Superconductivity, Project 93195.

\end {document}